\documentclass[aps,prl,preprint,tightenlines,superscriptaddress,showpacs,
byrevtex]{revtex4}
\usepackage{dcolumn}  

\usepackage{epsfig}
\newcommand{\mpp}{M_{p\bar{p}}}
\newcommand{\mb}{{M_{\rm bc}}}
\newcommand{\de}{{\Delta{E}}}
\newcommand{\bp}{{B^{+}}}

\newcommand{\pp}{{p\bar{p}}}
\newcommand{\ppk}{{p\bar{p}K^+}}

\newcommand{\ppks}{{p\bar{p}K_S^0}}

\newcommand{\plpi}{{p\bar{\Lambda}\pi^-}}

\newcommand{\ks}{{K_S^0}}

\begin{document}
\preprint{\vbox{ \hbox{   }
                  \hbox{BELLE-CONF-0423}
                  \hbox{ICHEP Abstract 11-0670}
}}
\title{ \quad\\[0.5cm]
\boldmath  Study of the Baryon-Antibaryon Low-Mass Enhancements in 
Charmless Three-body 
Baryonic $B$ Decays}


\affiliation{Aomori University, Aomori}
\affiliation{Budker Institute of Nuclear Physics, Novosibirsk}
\affiliation{Chiba University, Chiba}
\affiliation{Chonnam National University, Kwangju}
\affiliation{Chuo University, Tokyo}
\affiliation{University of Cincinnati, Cincinnati, Ohio 45221}
\affiliation{University of Frankfurt, Frankfurt}
\affiliation{Gyeongsang National University, Chinju}
\affiliation{University of Hawaii, Honolulu, Hawaii 96822}
\affiliation{High Energy Accelerator Research Organization (KEK), Tsukuba}
\affiliation{Hiroshima Institute of Technology, Hiroshima}
\affiliation{Institute of High Energy Physics, Chinese Academy of Sciences, Beijing}
\affiliation{Institute of High Energy Physics, Vienna}
\affiliation{Institute for Theoretical and Experimental Physics, Moscow}
\affiliation{J. Stefan Institute, Ljubljana}
\affiliation{Kanagawa University, Yokohama}
\affiliation{Korea University, Seoul}
\affiliation{Kyoto University, Kyoto}
\affiliation{Kyungpook National University, Taegu}
\affiliation{Swiss Federal Institute of Technology of Lausanne, EPFL, Lausanne}
\affiliation{University of Ljubljana, Ljubljana}
\affiliation{University of Maribor, Maribor}
\affiliation{University of Melbourne, Victoria}
\affiliation{Nagoya University, Nagoya}
\affiliation{Nara Women's University, Nara}
\affiliation{National Central University, Chung-li}
\affiliation{National Kaohsiung Normal University, Kaohsiung}
\affiliation{National United University, Miao Li}
\affiliation{Department of Physics, National Taiwan University, Taipei}
\affiliation{H. Niewodniczanski Institute of Nuclear Physics, Krakow}
\affiliation{Nihon Dental College, Niigata}
\affiliation{Niigata University, Niigata}
\affiliation{Osaka City University, Osaka}
\affiliation{Osaka University, Osaka}
\affiliation{Panjab University, Chandigarh}
\affiliation{Peking University, Beijing}
\affiliation{Princeton University, Princeton, New Jersey 08545}
\affiliation{RIKEN BNL Research Center, Upton, New York 11973}
\affiliation{Saga University, Saga}
\affiliation{University of Science and Technology of China, Hefei}
\affiliation{Seoul National University, Seoul}
\affiliation{Sungkyunkwan University, Suwon}
\affiliation{University of Sydney, Sydney NSW}
\affiliation{Tata Institute of Fundamental Research, Bombay}
\affiliation{Toho University, Funabashi}
\affiliation{Tohoku Gakuin University, Tagajo}
\affiliation{Tohoku University, Sendai}
\affiliation{Department of Physics, University of Tokyo, Tokyo}
\affiliation{Tokyo Institute of Technology, Tokyo}
\affiliation{Tokyo Metropolitan University, Tokyo}
\affiliation{Tokyo University of Agriculture and Technology, Tokyo}
\affiliation{Toyama National College of Maritime Technology, Toyama}
\affiliation{University of Tsukuba, Tsukuba}
\affiliation{Utkal University, Bhubaneswer}
\affiliation{Virginia Polytechnic Institute and State University, Blacksburg, Virginia 24061}
\affiliation{Yonsei University, Seoul}
  \author{K.~Abe}\affiliation{High Energy Accelerator Research Organization (KEK), Tsukuba} 
  \author{K.~Abe}\affiliation{Tohoku Gakuin University, Tagajo} 
  \author{N.~Abe}\affiliation{Tokyo Institute of Technology, Tokyo} 
  \author{I.~Adachi}\affiliation{High Energy Accelerator Research Organization (KEK), Tsukuba} 
  \author{H.~Aihara}\affiliation{Department of Physics, University of Tokyo, Tokyo} 
  \author{M.~Akatsu}\affiliation{Nagoya University, Nagoya} 
  \author{Y.~Asano}\affiliation{University of Tsukuba, Tsukuba} 
  \author{T.~Aso}\affiliation{Toyama National College of Maritime Technology, Toyama} 
  \author{V.~Aulchenko}\affiliation{Budker Institute of Nuclear Physics, Novosibirsk} 
  \author{T.~Aushev}\affiliation{Institute for Theoretical and Experimental Physics, Moscow} 
% ...

\collaboration{The Belle Collaboration}


\begin{abstract}
The angular distributions 
of the baryon-antibaryon low-mass enhancements seen in
the charmless three-body baryonic $B$ decays
$B^+ \to p \bar{p} K^+$,
$B^0 \to p \bar{p} K_S^0$, and $B^0 \to p \bar{\Lambda} \pi^-$ are reported. 
Searches for the
pentaquarks $\Theta^+$ and $\Theta^{++}$ in the relevant decay modes and
possible glueball states in the $p \bar{p}$ systems
are presented.  
The analysis is based on a
140~fb$^{-1}$ data sample recorded on the $\Upsilon({\rm 4S})$
resonance with the Belle detector at the KEKB asymmetric-energy $e^+e^-$
collider. 

\end{abstract}
\maketitle
\tighten
{\renewcommand{\thefootnote}{\fnsymbol{footnote}}
\setcounter{footnote}{0}

Observations of several baryonic three-body $B$ decays have been
reported recently~\cite{ppk,plpi,pph}.
One common feature of these observations is the peaking of
the baryon-antibaryon pair mass 
spectra toward threshold, as originally conjectured 
in Refs.~\cite{HS,rhopn} and elaborated more recently in
Refs.~\cite{glueball,RosnerB,rus}. The same peaking behavior
near threshold has been found in $J/\psi$ decays~\cite{BES} as well,
indicating that this might be a universal phenomenon.
The possible explanations include the presence of intermediate
gluonic states or side effects of the quark fragmentation process.
Alternatively, the dynamical picture can be replaced
by an effective range analysis with a baryon form factor. 
To distinguish among the above production
mechanism hypotheses, we study the threshold enhancements by examining the
angular distributions in the helicity frame for the
$\ppk$, $\ppks$ and $\plpi$~\cite{conjugate} modes.  Also, we update the mass
spectra from our previous studies. 

We use a  140 fb$^{-1}$  data sample,
consisting of 152 $ \times 10^6 B\bar{B}$ pairs,
collected by the Belle detector 
at the KEKB asymmetric energy $e^+e^-$ (3.5 on 8~GeV) collider~\cite{KEKB}.
The Belle detector is a large solid angle magnetic spectrometer
that consists of a three layer silicon vertex detector (SVD), a 50
layer central drift chamber (CDC), an array of aerogel threshold
\v{C}erenkov counters (ACC), a barrel-like arrangement of time of
flight scintillation counters (TOF), and an electromagnetic
calorimeter comprised of CsI(Tl) crystals located inside a
superconducting solenoid coil that provides a 1.5~T magnetic
field.  An iron flux return located outside of the coil is
instrumented to detect $K_L^0$ mesons and to identify muons. The
detector is described in detail elsewhere~\cite{Belle}.

The event selection criteria are based on the information obtained
from the tracking system
(SVD+CDC) and the hadron identification system (CDC+ACC+TOF).
All primary charged tracks
are required to satisfy track quality criteria
based on the track impact parameters relative to the
interaction point (IP). 
The deviations from the IP position are required to be within
$\pm$1 cm in the transverse ($x$--$y$) plane, and within $\pm$3 cm
in the $z$ direction, where the $z$ axis is opposite the
positron beam line. For each track, the likelihood values $L_p$,
$L_K$, and $L_\pi$ that it is a proton, kaon, or pion, respectively,
are determined from the information provided by
the hadron identification system.  The track is identified as a proton
if $L_p/(L_p+L_K)> 0.6 $ and $L_p/(L_p+L_{\pi})> 0.6$, or as a kaon if
$L_K/(L_K+L_{\pi})> 0.6$, or as a pion if $L_{\pi}/(L_K+L_{\pi})> 0.6$.
Candidate $\ks$ mesons
are reconstructed from pairs of oppositely charged tracks (both treated as
pions)
having an invariant mass consistent with  the $\ks$ nominal mass, 
as well as
a displaced vertex and flight direction consistent with
an origin at the IP.
A candidate $\Lambda$ baryon is reconstructed from a pair of oppositely
charged tracks---treated as a proton and negative pion---whose invariant mass
is consistent with the nominal $\Lambda$ baryon mass.  The proton-like
daughter is required to satisfy $L_p/(L_p+L_{\pi})> 0.6$.

Candidate $B$ mesons are reconstructed from the related final state
particles for the $B^+ \to p \bar{p} K^+$,
$B^0 \to p \bar{p} K_S^0$, and $B^0 \to p \bar{\Lambda} \pi^-$  modes.
We use two kinematic variables in the center of mass (CM) frame to identify the
reconstructed $B$ meson candidates: the beam energy
constrained mass $\mb = \sqrt{E^2_{\rm beam}-p^2_B}$, and the
energy difference $\de = E_B - E_{\rm beam}$, where $E_{\rm
beam}$ is the beam energy, and $p_B$ and $E_B$ are the momentum and
energy, respectively, of the reconstructed $B$ meson.
The fit region is
defined as 5.20 GeV/$c^2 < \mb < 5.29$ GeV/$c^2$ and -0.1 GeV $ < \de< 0.2$
GeV. From a GEANT based Monte Carlo (MC) simulation, the signal
peaks in the 
subregion 5.27 GeV/$c^2 < \mb < 5.29$ GeV/$c^2$ and $|\de|< 0.05$ GeV.
The lower bound of $\de$ is chosen to exclude possible contamination from
so-called ``cross-feed'' baryonic $B$ decays.

The background in the fit region solely arises from the continuum $e^+e^-
\to q\bar{q}$ ($q = u,\ d,\ s,\ c$) process.
We suppress the jet-like continuum background events relative to the more
spherical $B\bar{B}$ signal events using a Fisher discriminant~\cite{fisher}
that combines seven event shape variables, as described in Ref.~\cite{etapk}.
Probability density functions (PDFs) for the Fisher discriminant and
the cosine of the angle between the $B$ flight direction
and the beam direction in the $\Upsilon({\rm 4S})$ rest frame
are combined to form the signal (background)
likelihood ${\cal L}_{s}$ (${\cal L}_{b}$).
The signal PDFs are determined using signal MC
simulation; the background PDFs are obtained from 
the side-band data 
with $\mb < 5.26$ GeV/$c^2$.
We require
the likelihood ratio ${\cal R} = {\cal L}_s/({\cal L}_s+{\cal L}_b)$ 
to be greater than 0.7, 0.75, and 0.8 for the
$\ppk$, $\ppks$, and $\plpi$ modes, respectively.
These selection
criteria are determined by optimization of $n_s/\sqrt{n_s+n_b}$, where $n_s$ 
and $n_b$
denote the expected numbers of signal and background events, respectively. 
We use the branching fractions from our previous measurements~\cite{ppk,plpi,pph} in the calculation of $n_s$.
If there are  multiple $B$ candidates in one event, we 
select the one with the best $\chi^2$ value from the $B$ meson
vertex fit, in which
only the primary charged tracks 
are used.
Based on previous studies~\cite{ppk,plpi,pph}, 
we require the invariant mass of the baryon pair to be less than 
2.85 GeV/$c^2$ for the study of the threshold enhancement effect that
follows.

\begin{figure}[p]
\centering
\epsfig{file=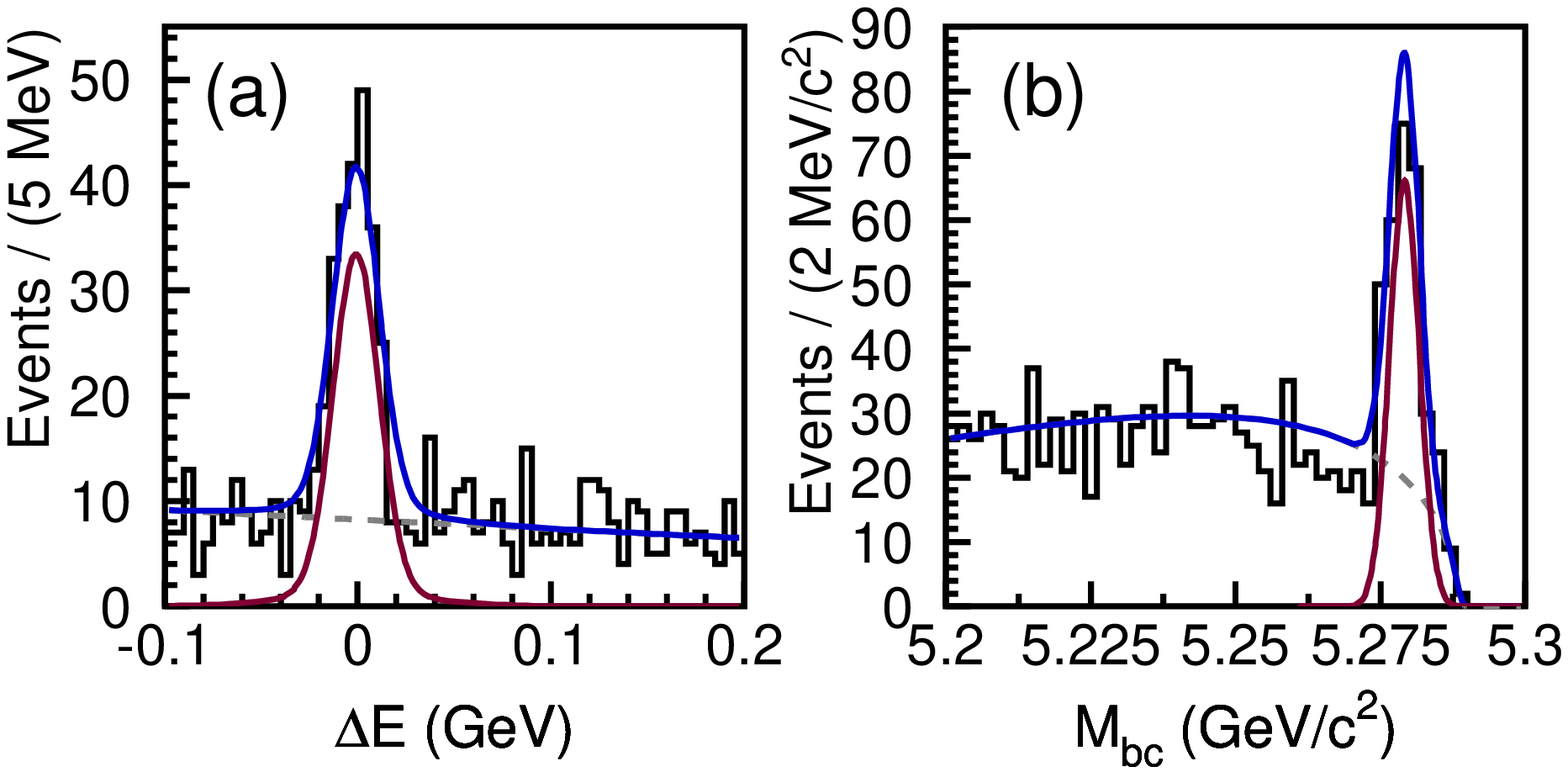,width=10cm}\\

\epsfig{file=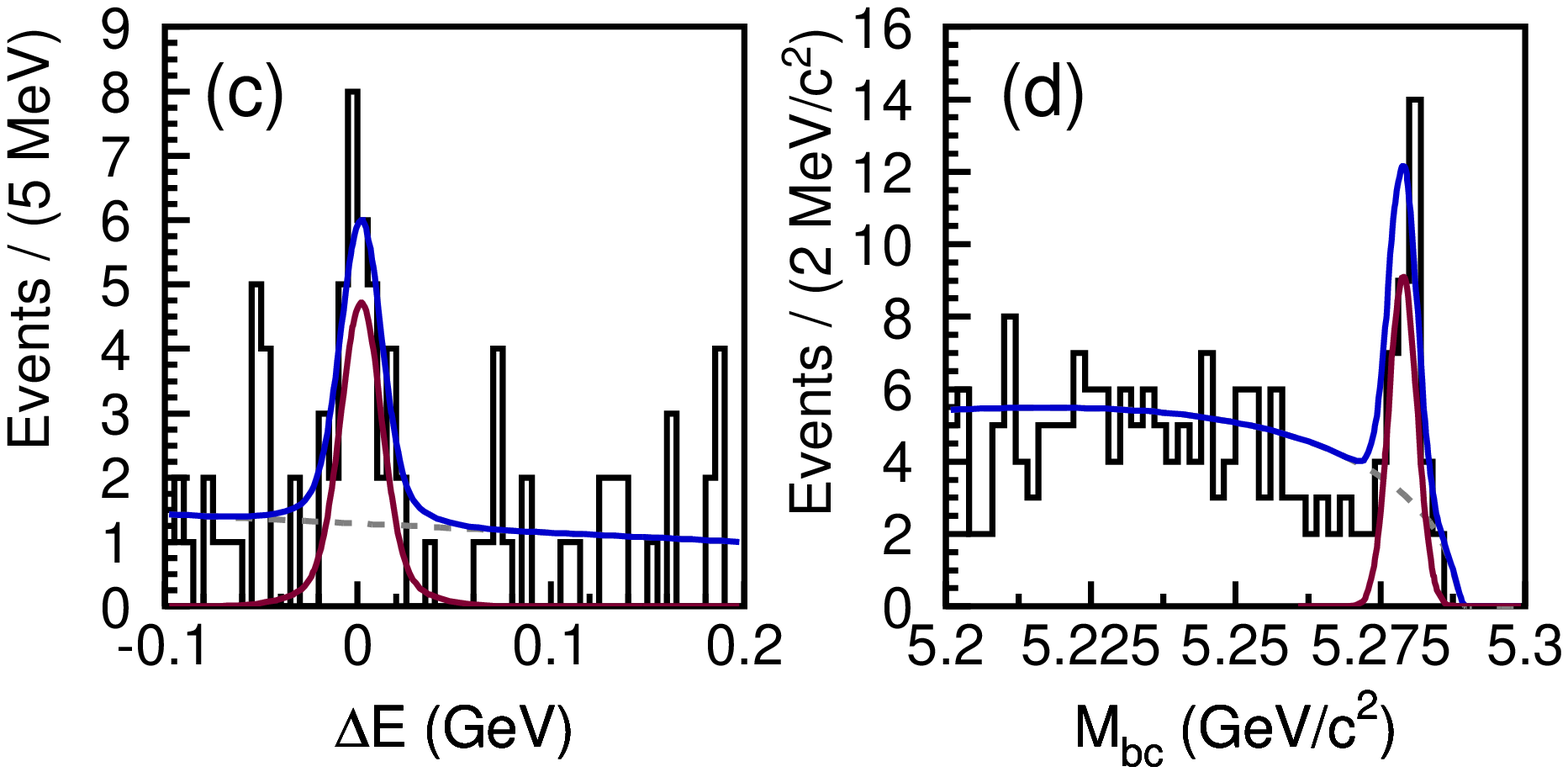,width=10cm}\\

\epsfig{file=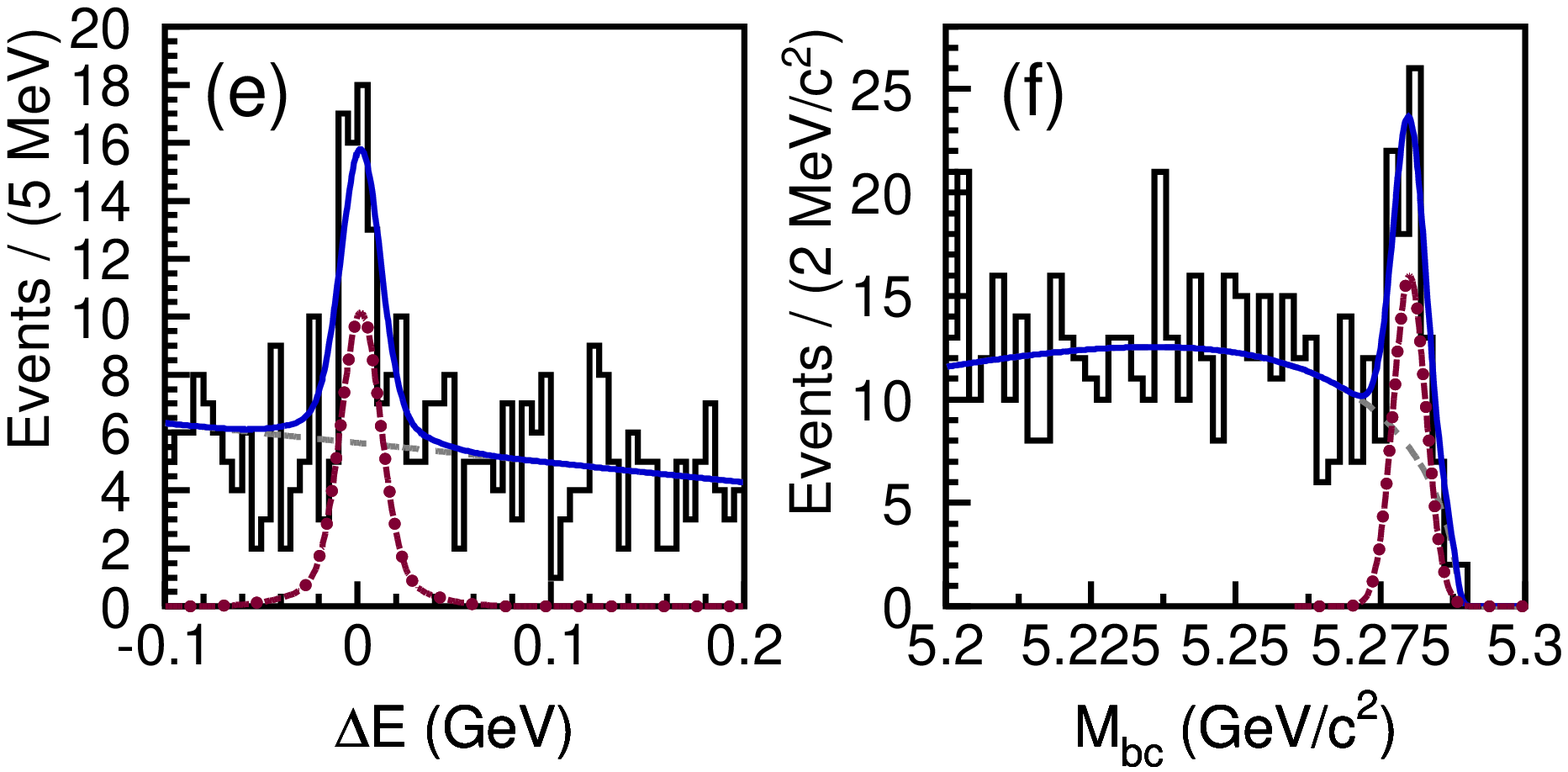,width=10cm}\\

\centering
\caption{ Distributions of $\de$ and $\mb$, respectively, for
(a) and (b) $\ppk$, (c) and (d) $\ppks$, and (e) and (f) $\plpi$ 
modes with baryon-antibaryon pair mass less than  
2.85 GeV/$c^2$. 
The blue, red and dashed lines represent the combined fit
result, fitted signal and fitted background, respectively.
}

\label{fg:mergembde}
\end{figure}

The $\mb$ distributions (with $|\de|<$ 0.05
GeV), and the $\de$ distributions (with $\mb >$ 5.27 GeV/$c^2$)
for the $\ppk$, $\ppks$ and $\plpi$ modes are
shown in Fig.~\ref{fg:mergembde}. 
We use an unbinned likelihood fit that maximize the likelihood function, 
$$ L = {e^{-(N_s+N_b)} \over N!}\prod_{i=1}^{N} 
\left[\mathstrut^{\mathstrut}_{\mathstrut}N_s P_s(M_{{\rm bc}_i},\Delta{E}_i)+
N_b P_b(M_{{\rm bc}_i},\Delta{E}_i)\right],$$
to estimate the signal yield;
here $P_s\ (P_b)$ denotes the signal (background) PDF, 
$N$ is the number of events in the fit, and $N_s$ and $N_b$
are free parameters representing the number of signal and background
events, respectively.

For the signal PDF,
we use the product of a Gaussian in $\mb$ and a double Gaussian in $\de$.
We fix
the parameters of these functions to values determined by MC simulation
~\cite{correction}.
The continuum background PDF 
is taken as the product of shapes in
$\mb$ and $\de$, which are assumed to be uncorrelated.
These shapes are obtained from sideband
events, with 0.1 GeV $ < \de < 0.2$ GeV for the $\mb$ function and
with 5.20 GeV/$c^2$ $ < \mb <$ 5.26 GeV/$c^2$ for the $\de$ function,
and are confirmed with a continuum MC sample.
We use the following parametrization first used by 
the ARGUS collaboration~\cite{Argus}, 
$ f(\mb)\propto \mb\sqrt{1-x^2}
\exp[-\xi (1-x^2)]$,  
to model
the $\mb$ background, with $x$ given by $\mb/E_{\rm beam}$ and $\xi$ as
a fit parameter. 
The $\de$ background shape is modeled by a first order polynomial whose slope
is a fit parameter.
The projections of the fit results are shown in 
Fig.~\ref{fg:mergembde} by solid curves. 
The fit yields are
$217 \pm 17$,
28.6 $^{+6.5}_{-5.8}$,
and 48.8 $^{+8.2}_{-7.5}$
for the $\ppk$, $\ppks$, and $\plpi$ modes, respectively. 

\begin{figure}[b!]
\centering
\epsfig{file=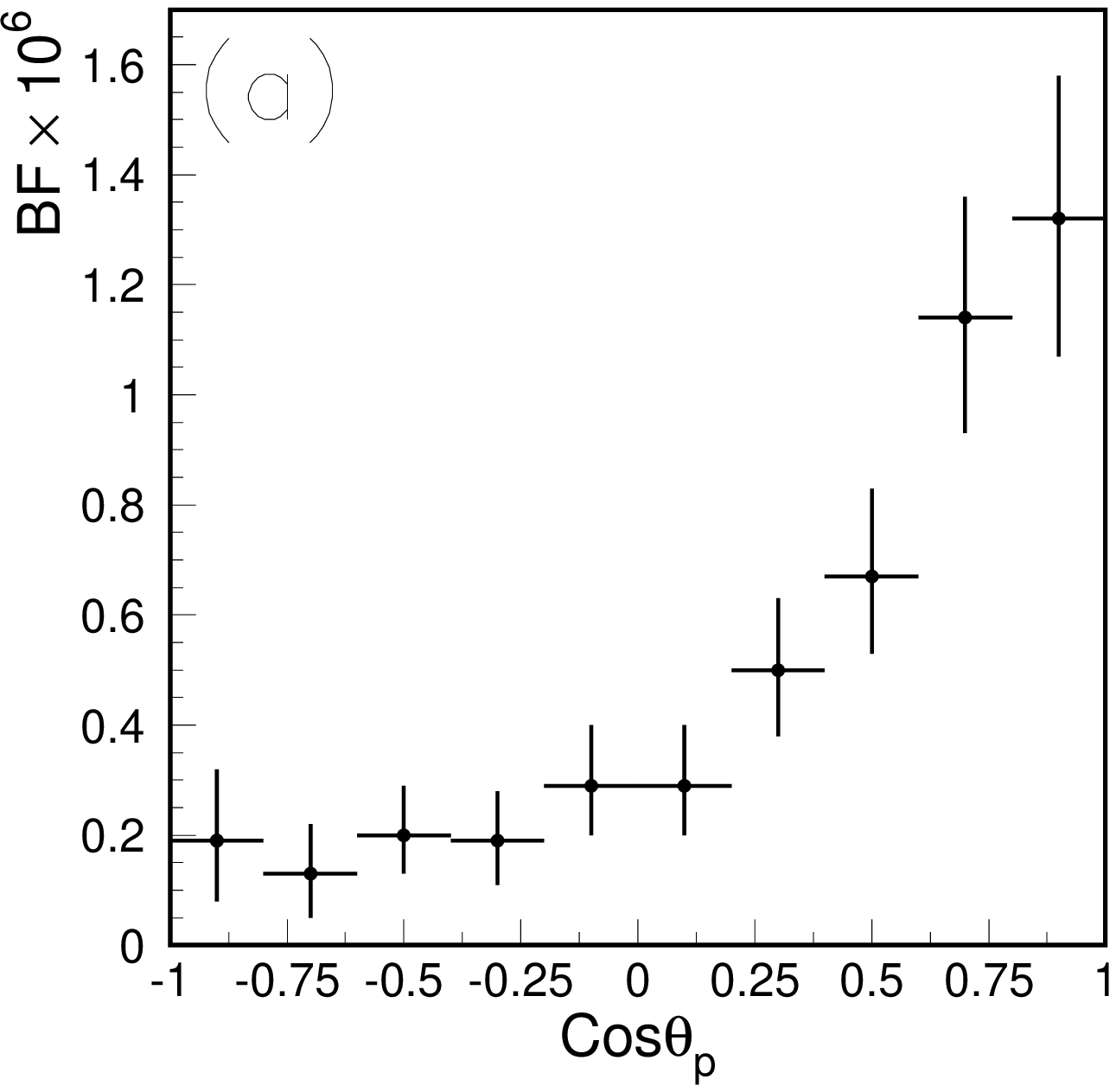,width=5cm}\enspace%
\epsfig{file=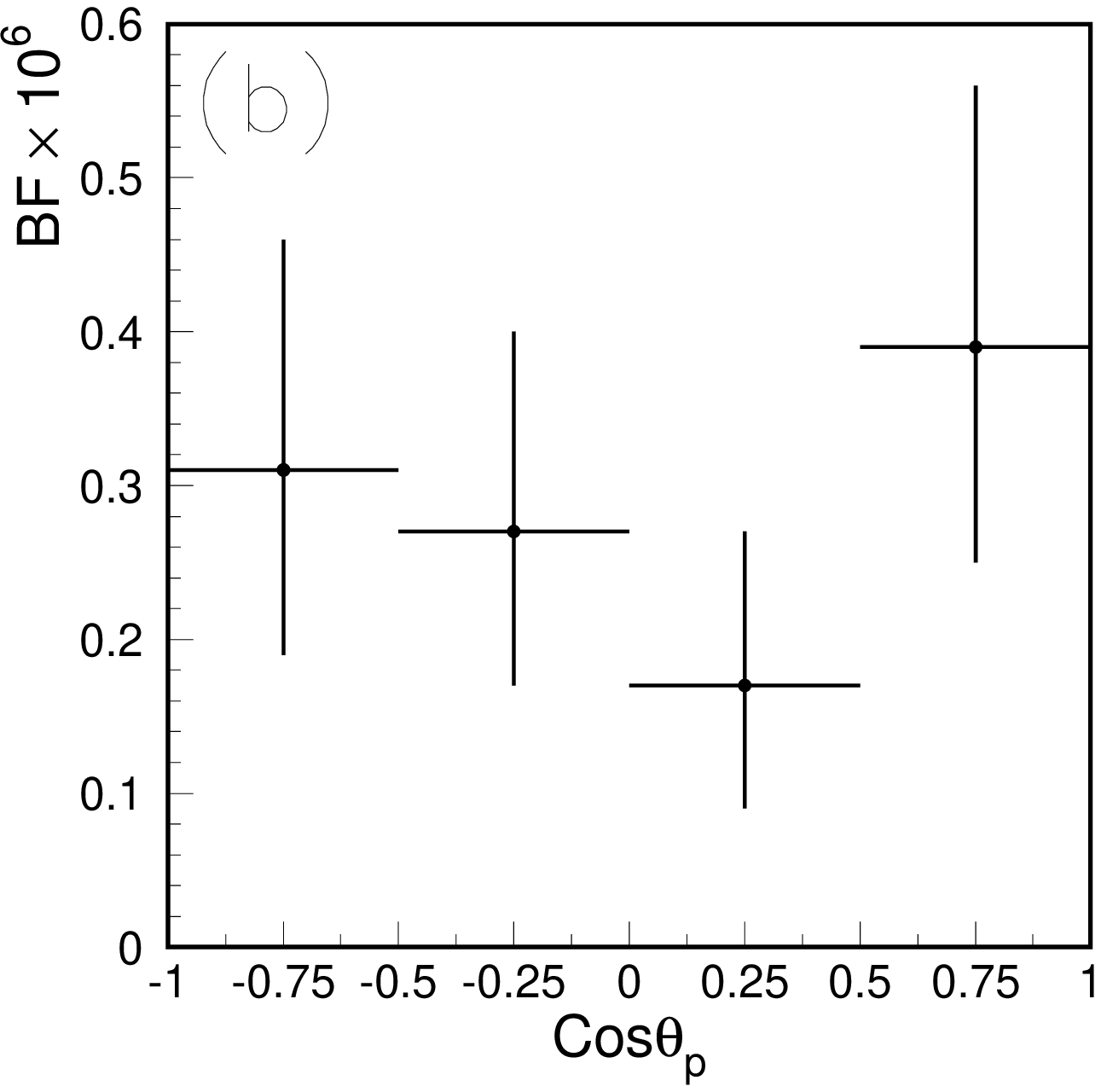,width=5cm}\enspace%
\epsfig{file=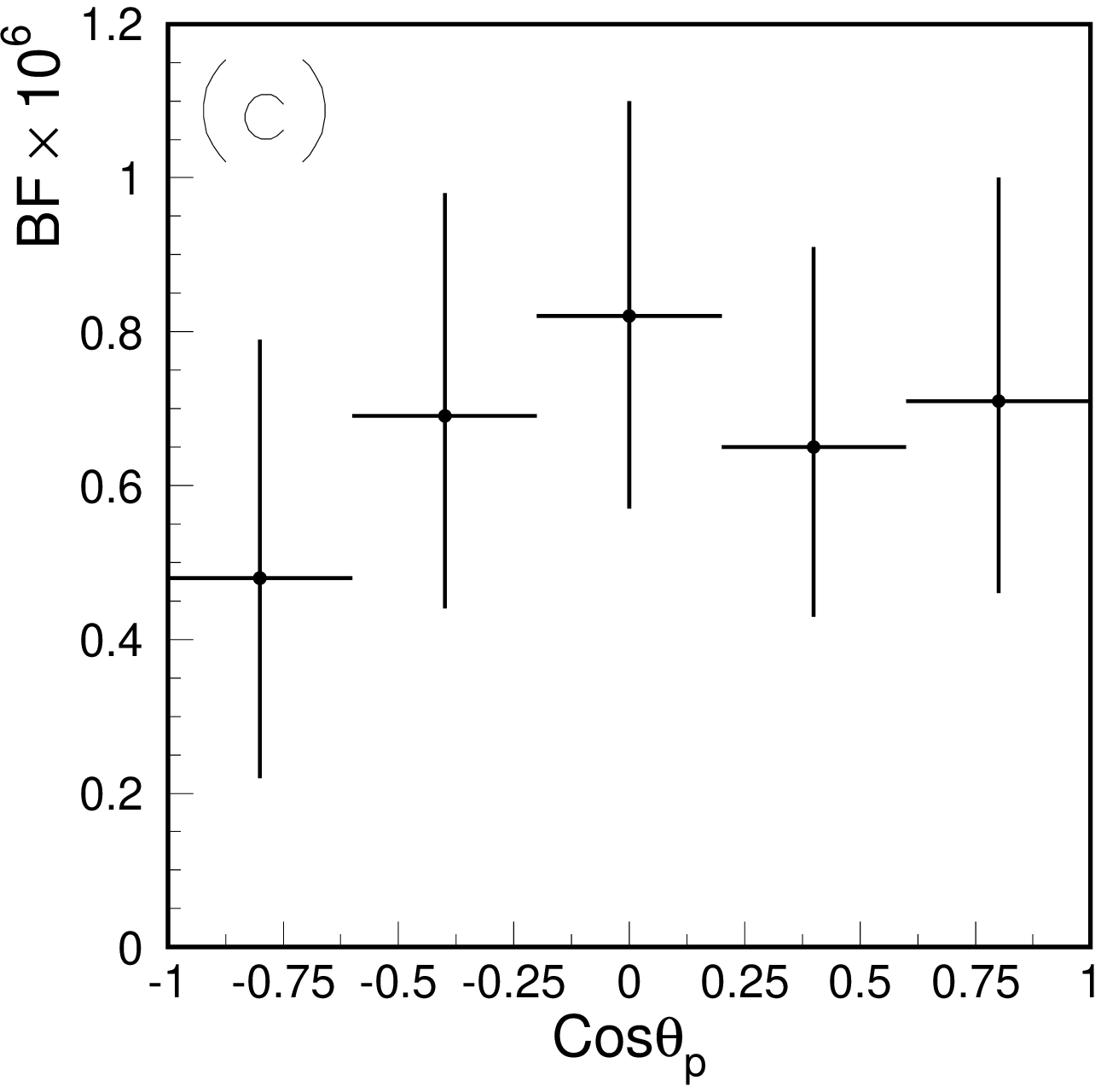,width=5cm}\\
\epsfig{file=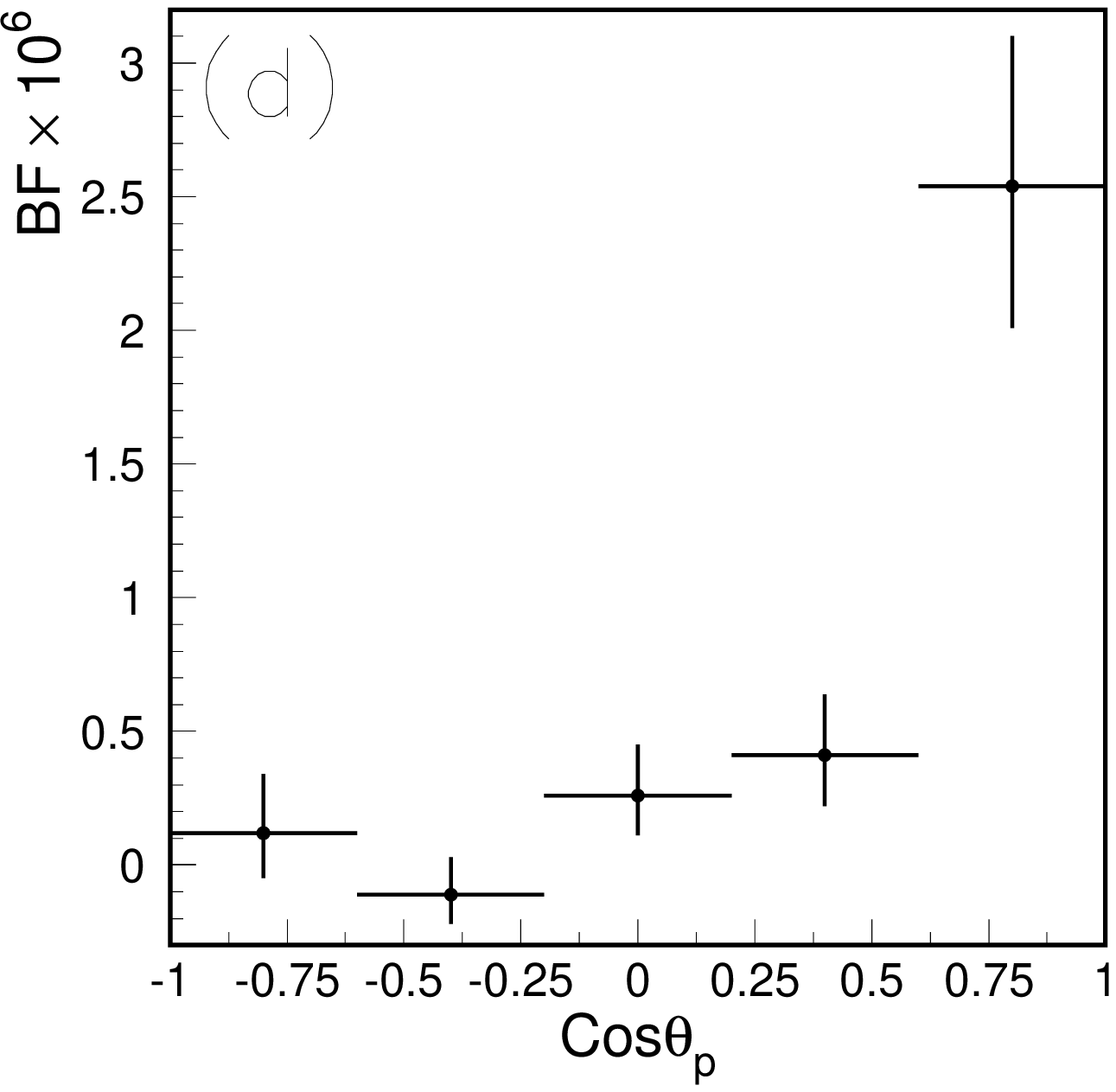,width=5cm}\enspace%
\epsfig{file=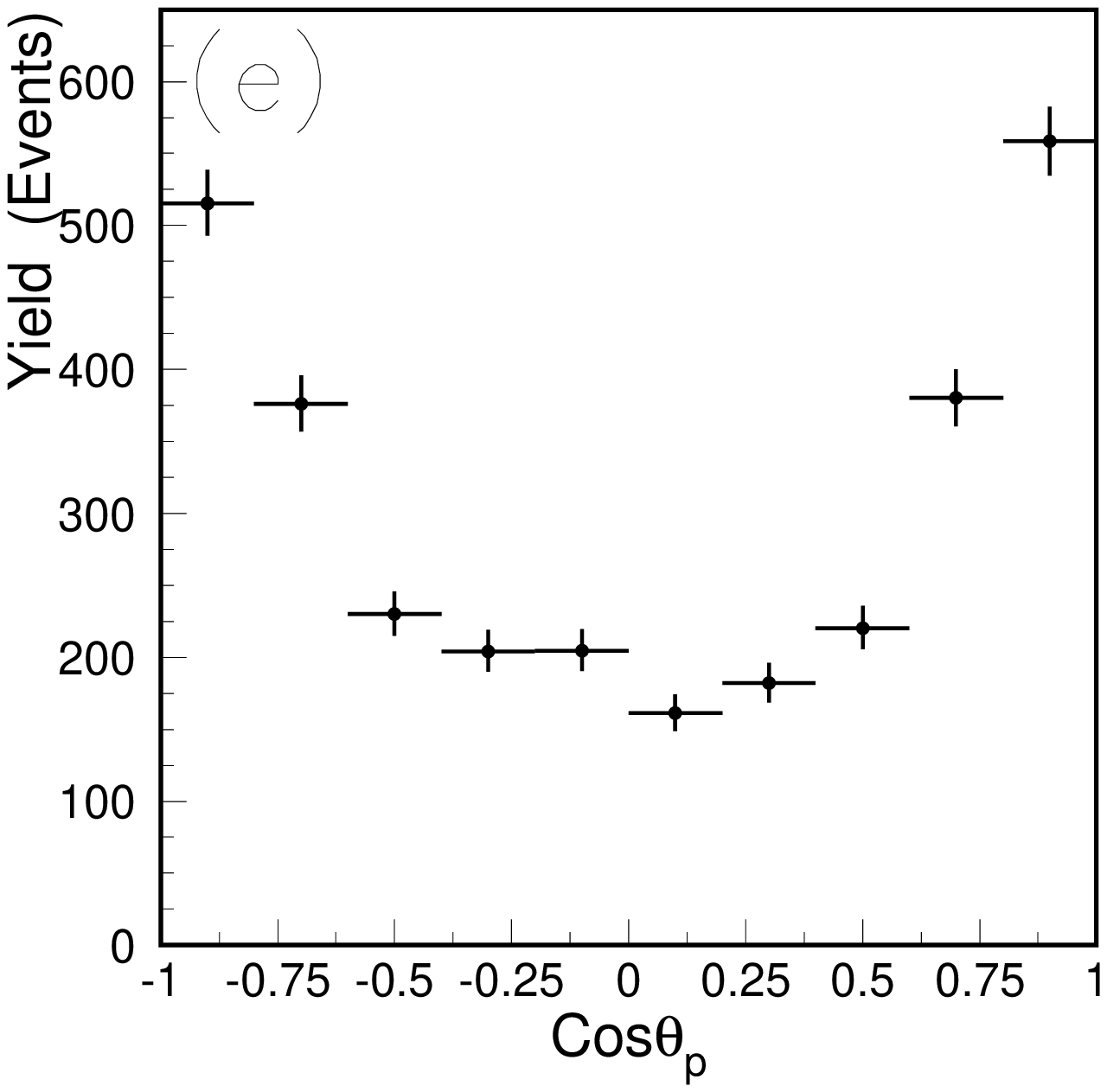,width=5cm}\enspace%
\epsfig{file=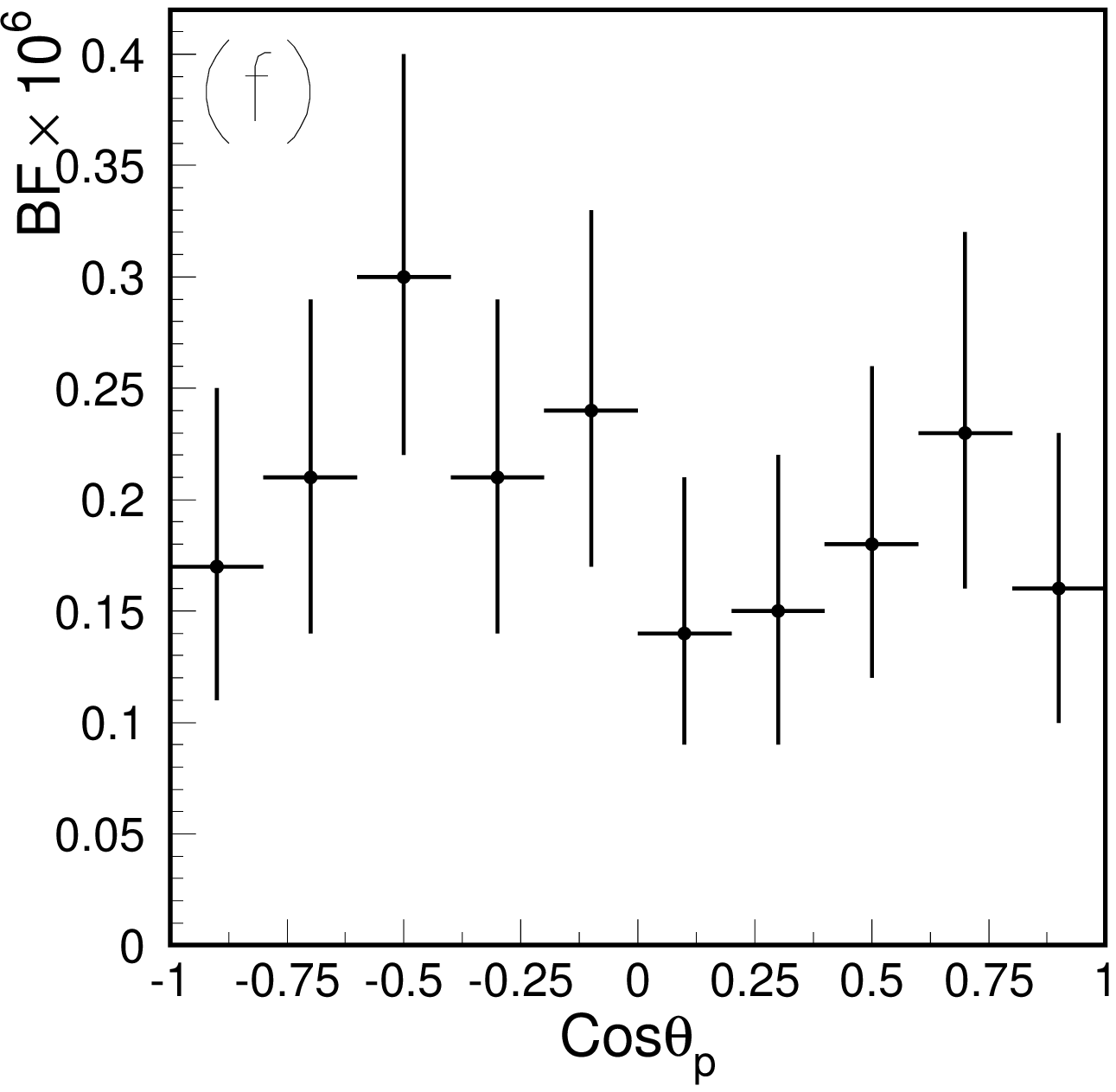,width=5cm}\\
\centering
\caption{Branching fraction {\it vs.}
proton helicity angle in the baryon-antibaryon pair system for
(a) $\ppk$, (b) $\ppks$, and (c) $\plpi$ modes. (d) The proton angular
distribution of the $p \pi^-$ system 
against the $\bar{\Lambda}$ direction 
in the $\plpi$ mode.
(e) Background yield of $\ppk$ in the fit. 
(f) Comparison with the
$J/\psi$ mass region. 
}
\label{fg:thetap}
\end{figure}

We study the proton angular distribution of the baryon-antibaryon pair system
in its helicity frame. The angle $\theta_p$ is defined as the
angle between the proton direction and the opposite
of the light meson direction in the baryon-antibaryon pair rest frame.  
Note that after charge conjugation, 
the angle is determined by $\bar{p}$ and $K^+$ 
(or $p$ and $K^-$) 
for the $\ppk$ mode. Fig.~\ref{fg:thetap}(a)-(c) shows the
branching fractions as a function of $\cos \theta_p$. The error bars
include the statistical uncertainty from the fit
and the systematic uncertainty. It is clear that the fragmentation
process is favored for the $\ppk$ mode. Protons are emitted along
the $K^-$ direction most of the time, which can be explained by 
a parent $b \to s$ penguin transition followed by $s\bar{u}$ fragmentation into the
final state. The $\cos \theta_p$ distribution of the $\ppks$ mode does not
have enough statistics to support or refute this interpretation, 
although it seems to be
peaked towards $\pm 1$
since the flavor information is not applied in this case.
The distribution for the $\plpi$ mode 
is quite flat, in contrast to that of the
$\ppk$ mode, although both presumably share a common origin in the
$b \to s$ transition. 
In fact, this parentage suggests that it would be more suitable to draw
the proton angular distribution of the $p \pi^-$ pair
relative to the $\bar{\Lambda}$ 
direction; this is shown in Fig.~\ref{fg:thetap}(d). It is evident 
that the above interpretation is supported: 
the proton tends to emerge parallel to the $\bar{\Lambda}$ baryon.
As a cross check, the distribution of $\cos \theta_p$
for background events in the $\ppk$ sample
is shown in Fig.~\ref{fg:thetap}(e). (Similar distributions are obtained 
for the backgrounds of the $\ppks$ and $\plpi$ modes.) The background has
a $1+\cos^2 \theta_p $ distribution, which can be explained  as
arising from the random combination of two high momentum particles from the
$q \bar{q}$ jets.  The fragmentation signature is not seen in
the $B^+ \to J/\psi K^+$ mode, where the $J/\psi$ meson decays to a
$\pp$ pair.
For $J/\psi$ candidates with invariant mass in the range
$3.07$ GeV/$c^2 < \mpp < 3.11$ GeV/$c^2$,
the distribution of $\cos \theta_p$ is flat,
as shown in Fig.~\ref{fg:thetap}(f).
 
\begin{figure}[b!]
\centering
\epsfig{file=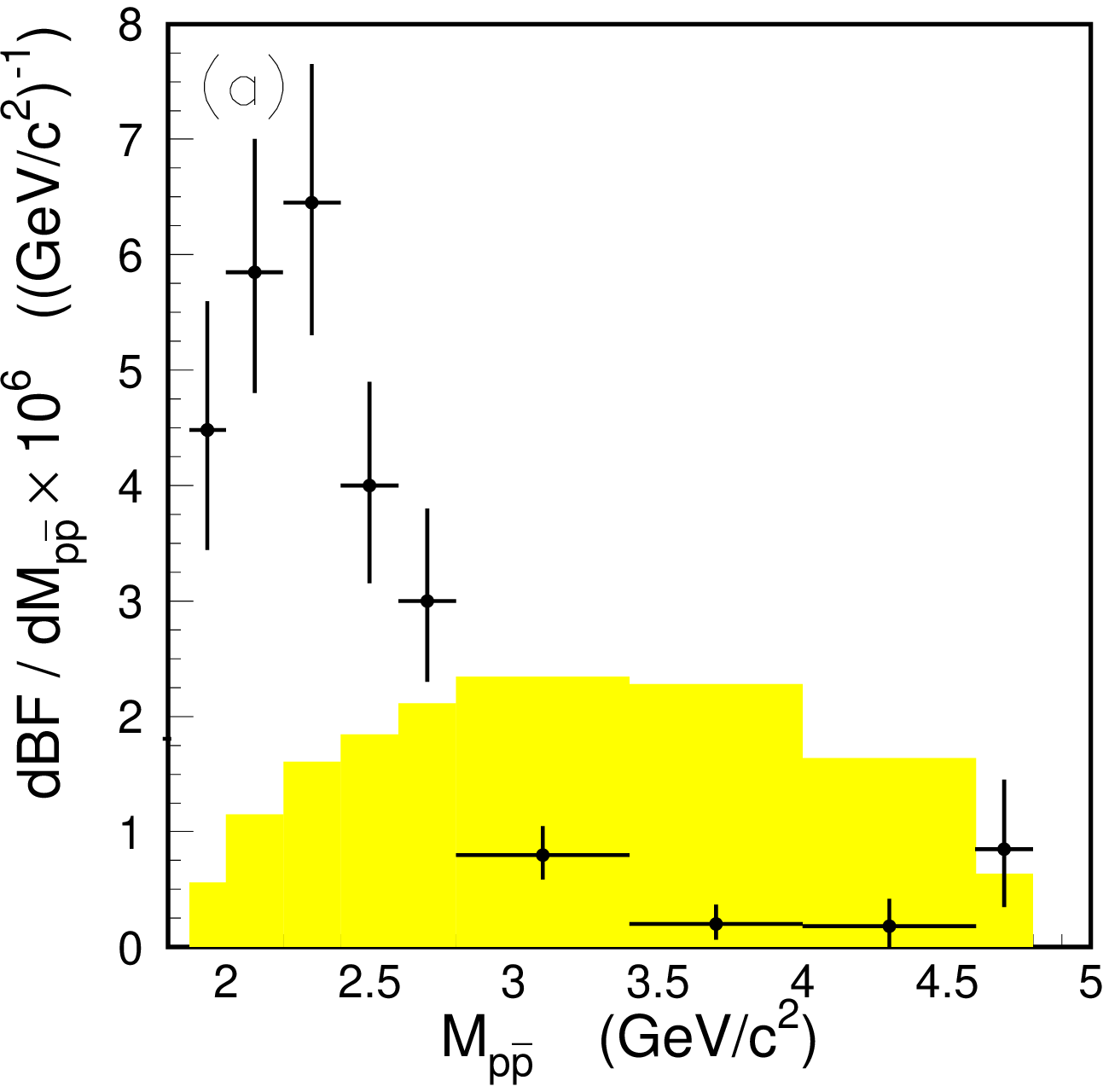,width=7cm}\\
\epsfig{file=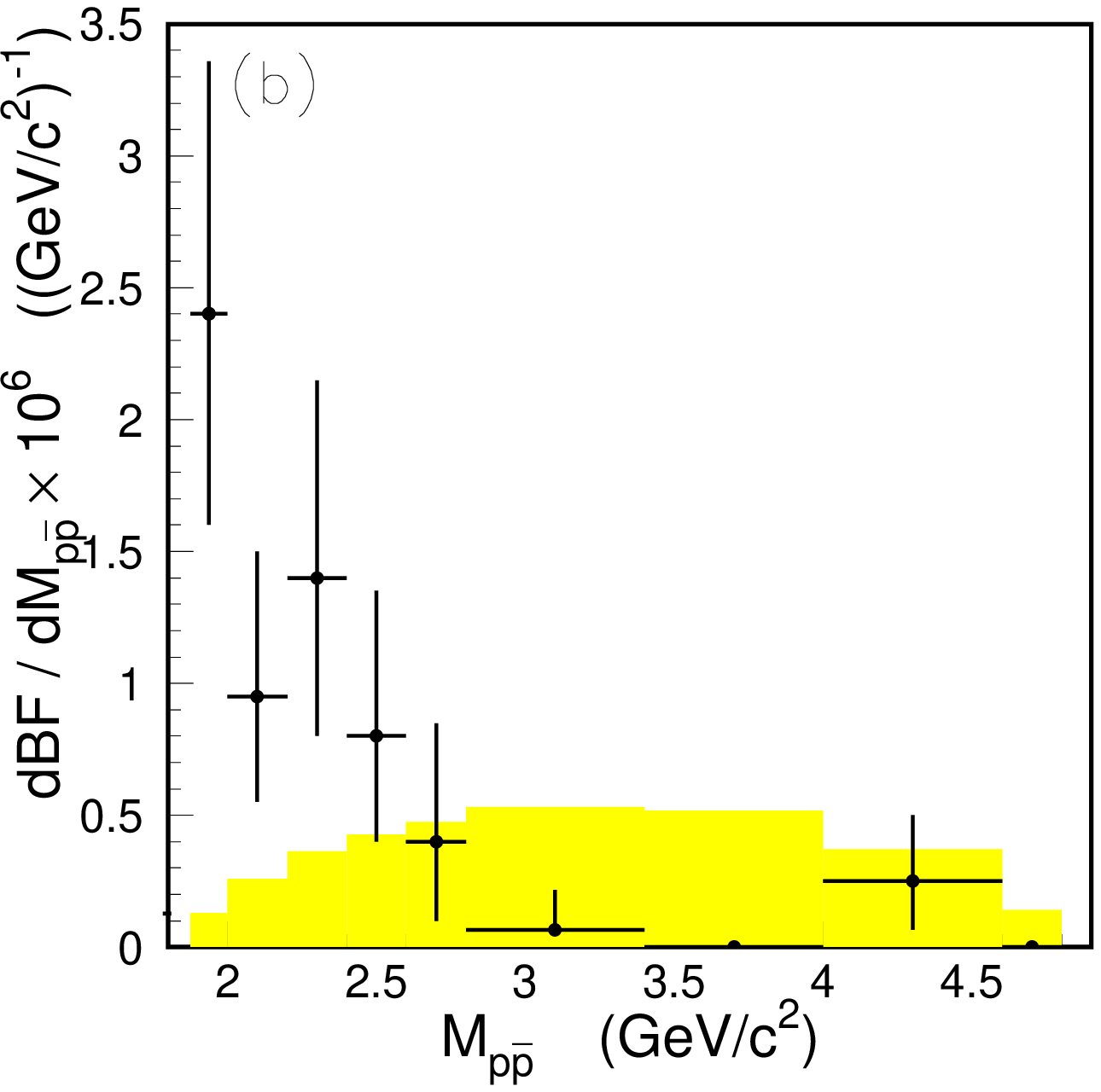,width=7cm}\\
\epsfig{file=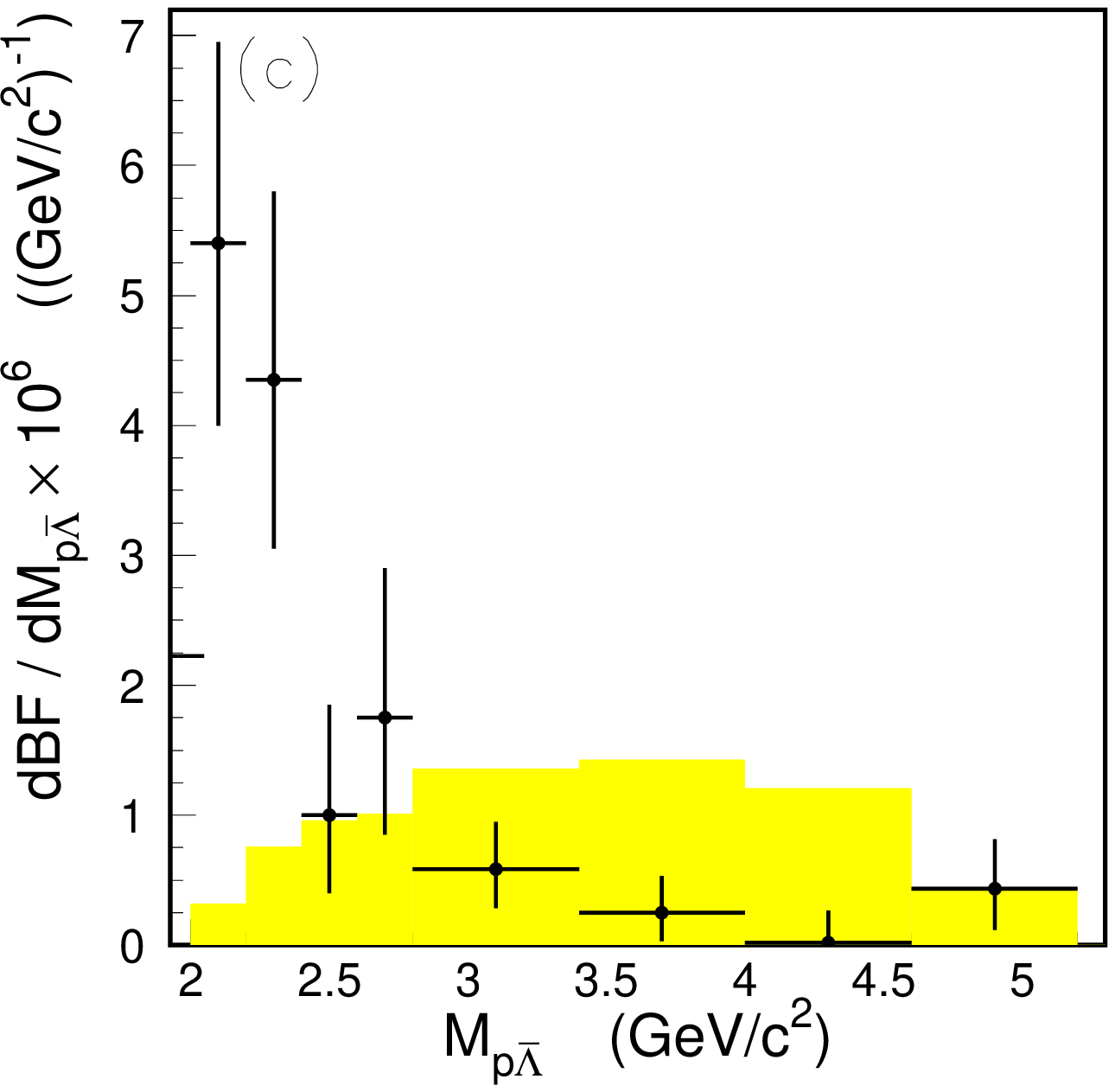,width=7cm}\\

\centering
\caption{Differential branching fraction for 
(a) $\ppk$, (b) $\ppks$, and (c) $\plpi$ 
modes 
as a function of baryon-antibaryon pair mass. The shaded distribution
shows the expectation from a
phase-space MC simulation with area scaled to the
signal yield. A charmonium veto has been applied in (a) and (b).
}
\label{fg:allphase}
\end{figure}

The differential branching fraction as a function of the 
baryon pair invariant mass
is shown in Fig.~\ref{fg:allphase}.  Here, 
the efficiency as a function of baryon pair mass for each signal mode is
determined by MC simulation, with the events distributed uniformly
in phase space.
The regions  $2.850$ GeV/$c^2  <M_{p \bar{p}}<3.128$ GeV/$c^2$
and $3.315$ GeV/$c^2 <M_{p \bar{p}}<3.735$ GeV/$c^2$
are excluded to remove background from $B$ decay modes
containing an $\eta_c$, $J/\psi$,
$\psi^{\prime}$, $\chi_{c0}$, or $\chi_{c1}$ meson.
The width of the low mass enhancement in each distribution of
Fig.~\ref{fg:allphase} depends on the signal mode.
A different narrow width is seen also in the newly discovered
$\bp \to \Lambda \bar{\Lambda} K^+$ decay~\cite{LLK}.

Systematic uncertainties 
are determined using high statistics control data samples. For proton
identification, we use a  $\Lambda \to p \pi^-$ sample, while for
$K/\pi$ identification we use a $D^{*+} \to D^0\pi^+$,
 $D^0 \to K^-\pi^+$ sample.
Tracking efficiency is measured with
fully and partially reconstructed $D^*$ samples.
The $\ks$ reconstruction efficiency is determined from a $D^- \to \ks\pi^-$
sample. The $\Lambda$ and $\ks$ reconstruction efficiencies have the same
uncertainty due to off-IP tracks if the uncertainty of the daughter
proton identification criterion is not taken into account.  
The $\cal R$ continuum suppression uncertainty is estimated from
$b \to c$ control samples with similar final states.
Based on these studies,
we assign a 1\% error for each track, 3\% for each proton identification,
2\% for each kaon/pion identification, 5\% for $\ks$ and $\Lambda$
off-IP reconstruction
and 6\% for the $\cal R$ selection.

A systematic uncertainty of 4\% in the fit yield is determined by varying
the parameters of the signal and background PDFs.  
The MC statistical
uncertainty and binning of the baryon pair mass contribute a 2\%
error in the branching fraction determination. The error on the
number of $B\bar{B}$ pairs is 1\%, where the
assumption is made that the branching fractions of $\Upsilon({\rm 4S})$ 
to neutral and charged $B\bar{B}$ pairs are equal. 

We first sum the correlated errors linearly and then combine with the
uncorrelated ones in quadrature. The total systematic
uncertainties are 11\%, 12\%, and 12\% for
the $\ppk$, $\ppks$, and $\plpi$ modes,
respectively.

The newly observed narrow pentaquark state, $\Theta^+$~\cite{penta}, can
decay into $p \ks$. We perform a search in our data sample
by requiring   $1.53$ GeV/$c^2 < M_{p\ks} < 1.55 $ GeV/$c^2$.
The $\mb$ and $\de$ projection plots in Fig.~\ref{fg:penta} show
no evidence for a pentaquark signal. Since there are few events in the fit
 window, we fix the
background shapes from side-band data.
We use the fit results to estimate the expected background
and compare this  with the observed number of events
in the signal region
to set the upper limit on the
yield~\cite{Gary,Conrad}.
The systematic uncertainty is included in this estimation.
The upper limit yield is determined to be 3.9 at the 90\% confidence level. The
related upper limit product of branching fractions is 
${\cal B}(B^0 \to {\Theta^+}\bar{p})\times {\cal B}({\Theta^+
}\to p\ks) < 2.3 \times 10^{-7}$ at the 90\% confidence level.
We also perform a search for $\Theta^{++}$, which can decay to $p K^+$.
Because there are only theoretical conjectures for the existence of 
such a state, we examine the wider mass region of
$1.6$ GeV/$c^2 < M_{p K^+} < 1.8 $ GeV/$c^2$.  We
find no evidence for signal. Assuming this state is narrow and 
centered near 1.71 GeV/$c^2$, the
upper limit yield is 3.3 events at the 90\% confidence level. The
related upper limit product of branching fractions is
${\cal B}(\bp \to {\Theta^{++}}\bar{p})\times {\cal B}({\Theta^{++}
}\to p K^+) < 9.1 \times 10^{-8}$ at the 90\% confidence level.

\begin{figure}[p]
\centering
\epsfig{file=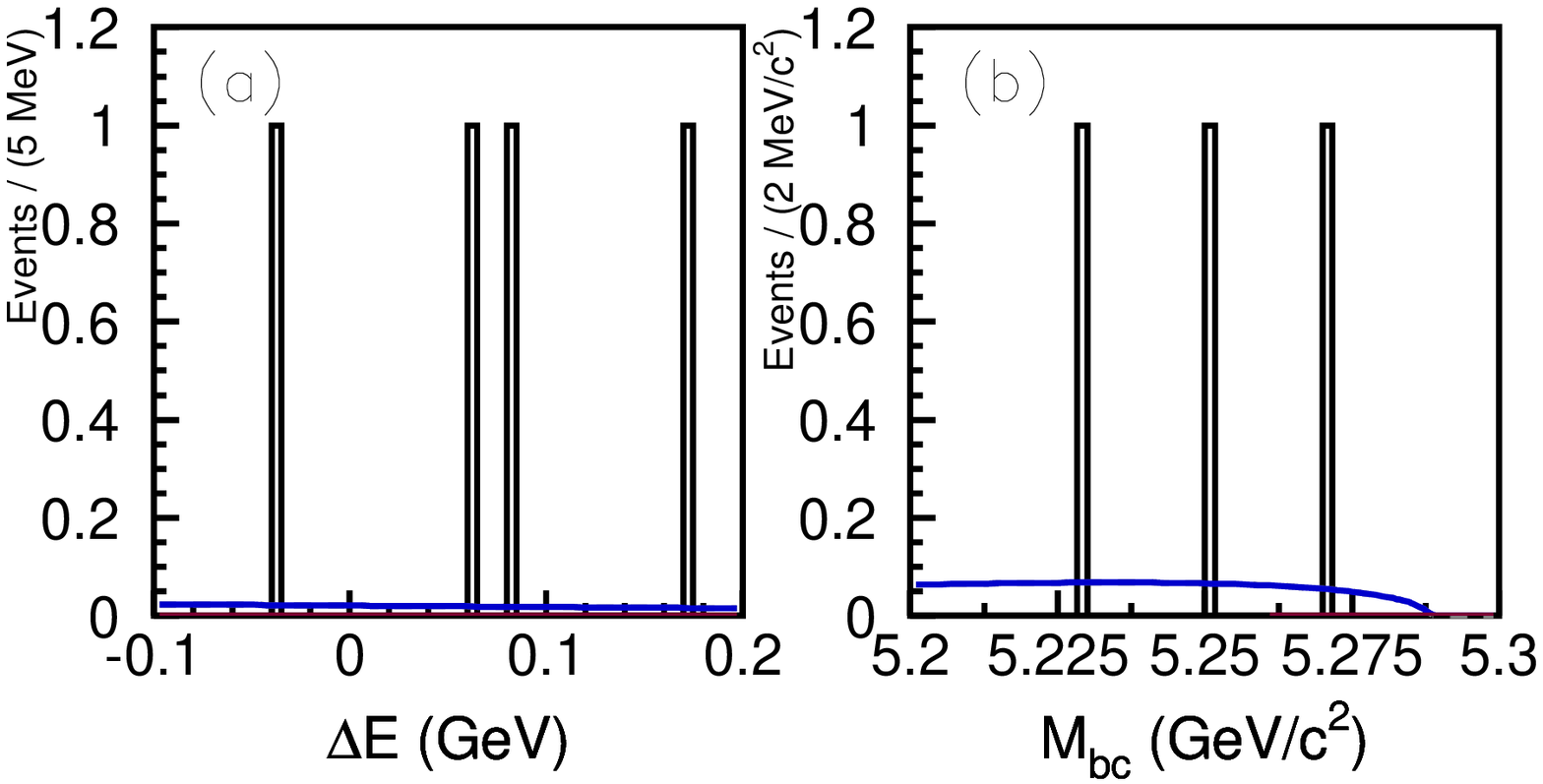,width=10cm}\\
\centering
\caption{Distributions of $\mb$ and $\de$ for the $\ppks$ mode
with $1.53$ GeV/$c^2 < M_{p\ks} < 1.55 $ GeV/$c^2$.
The curves represent the fit
projections. 
}

\label{fg:penta}
\end{figure}

One theoretical
conjecture~\cite{glueball} suggests that a glueball resonance with mass
near 2.3
GeV/$c^2$ may explain the $\mpp$ threshold
peaking behavior for the $\ppk$ mode. Since the $\mpp$ mass
resolution is about 10 MeV/$c^2$, we scan through the
$2.2$ GeV/$c^2 < \mpp < 2.4 $ GeV/$c^2$ mass region
with a 20 MeV/$c^2$ wide window. The highest upper 
limit yield is found to be 18.9. 
We use this data set to set an upper limit on the product of branching
fractions of
${\cal B}(\bp \to {\rm glueball}\ K^+)\times {\cal B}({\rm
glueball}\ \to \pp) < 4.1 \times 10^{-7}$ at the 90\%
confidence level for a possible narrow glueball state with  mass
in the 2.2 -- 2.4 GeV/$c^2$ range. The theoretical prediction is
around $1 \times 10^{-6}$.

In summary, using 152 $ \times 10^6 B\bar{B}$ events, we measure the
angular and invariant mass distributions of the baryon-antibaryon pair
system near threshold for the $\ppk$, $\ppks$ and $\plpi$  
baryonic $B$ decay modes. The quark fragmentation
process is supported, but the gluonic picture is disfavored. 
Searches for a $B$ meson decaying into pentaquark $\Theta^+$ or a
glueball in the above related modes give null results. We set stringent
upper limits on the product of the decay branching fractions. 

We thank the KEKB group for the excellent operation of the
accelerator, the KEK Cryogenics group for the efficient
operation of the solenoid, and the KEK computer group and
the National Institute of Informatics for valuable computing
and Super-SINET network support. We acknowledge support from
the Ministry of Education, Culture, Sports, Science, and
Technology of Japan and the Japan Society for the Promotion
of Science; the Australian Research Council and the
Australian Department of Education, Science and Training;
the National Science Foundation of China under contract
No.~10175071; the Department of Science and Technology of
India; the BK21 program of the Ministry of Education of
Korea and the CHEP SRC program of the Korea Science and
Engineering Foundation; the Polish State Committee for
Scientific Research under contract No.~2P03B 01324; the
Ministry of Science and Technology of the Russian
Federation; the Ministry of Education, Science and Sport of
the Republic of Slovenia; the National Science Council and
the Ministry of Education of Taiwan; and the U.S.\
Department of Energy.


\begin{thebibliography}{99}
\bibitem{conjugate} {Throughout this report,
inclusion of charge conjugate mode is always implied unless
otherwise stated.}
\bibitem{correction}{ There are small corrections applied to these
parameters based on the 
study of the difference between data and MC of $B \to D\pi$ decays.}
\end{thebibliography}
\end{document}